\documentclass[prl,twocolumn,showpacs,amsmath,amssymb]{revtex4}

\usepackage{epsfig}
\usepackage{dcolumn}
\usepackage{bm}

\begin{document}


\title{An Efficient Source of Single Photons: \\
    A Single Quantum Dot in a Micropost Microcavity}

\author{Matthew Pelton}
 \altaffiliation[Current Address: ]{Laboratory of Quantum Optics and Quantum Electronics,
 Department of Microelectronics and Information Technology, Royal Institute of Technology (KTH),
 Electrum 229, SE-164 40  Kista, Sweden}
 \email{pelton@imit.kth.se}
\author{Charles Santori}
\author{Jelena Vu\v{c}kovi\'{c}}
\author{Bingyang Zhang}
\altaffiliation[Also at ]{NTT Basic Research Laboratories, Atsugishi, Kanagawa, Japan}
\author{Glenn S. Solomon}
 \altaffiliation[Also at ]{Solid-State Photonics Laboratory, Stanford
 University, Stanford, California, 94305 U.S.A.}
\author{Jocelyn Plant}
\author{Yoshihisa Yamamoto}
\altaffiliation[Also at ]{NTT Basic Research Laboratories, Atsugishi, Kanagawa, Japan}
\affiliation{Quantum Entanglement Project, ICORP, JST,
         E. L. Ginzton Laboratory, Stanford University,
         Stanford, California, 94305 U.S.A.}

\date{\today}

\begin{abstract}
We have demonstrated efficient production of triggered single photons by coupling a single semiconductor quantum dot to a
three-dimensionally confined optical mode in a micropost microcavity.  The efficiency of emitting single photons into a
single-mode travelling wave is approximately 38\%, which is nearly two orders of magnitude higher than for a quantum dot in bulk
semiconductor material. At the same time, the probability of having more than one photon in a given pulse is reduced by a factor
of seven as compared to light with Poissonian photon statistics.
\end{abstract}

\pacs{42.50.Ct, 42.50.Dv, 78.67.Hc, 85.60.Jb}

\maketitle

The photon statistics of a light source can be described by the second-order autocorrelation function, defined as follows:
$g^{(2)}(\tau) = \langle \hat{a}^{\dag}(t)\hat{a}^{\dag}(t+\tau)\hat{a}(t+\tau)\hat{a}(t) \rangle / \langle \hat{a}^{\dag}\hat{a}
\rangle ^2$, where $\hat{a}^{\dag}(t)$ and $\hat{a}(t)$ are the photon creation and annihilation operators, respectively, at time
$t$.  A pulsed source will have a correlation function consisting of a series of peaks separated by the repetition period $T$. The
area $g^{(2)}_o$ of the peak around $\tau = 0$, normalized by $T$, gives an upper bound on the probability that two or more
photons are present in the same pulse: $ P(n \geq 2) \leq (1/2) \left \langle \hat{n} \right \rangle^2g^{(2)}_o$, where $\langle
\hat{n} \rangle$ is the mean photon number per pulse \cite{ref:Santori01}.  A source where $g^{(2)}_o < 1$ has a reduced
multi-photon probability as compared to coherent light with Poissonian photon statistics.  If $g^{(2)}_o$ is sufficiently close to
zero, we can speak of a {\em single-photon source}.

Such a source has been demonstrated using the controlled excitation of single molecules \cite{ref:Brunel99,ref:Lounis00} and
single nitrogen-vacancy centers in diamond nanocrystals \cite{ref:Beveratos02}, and using the controlled injection of carriers
into a mesoscopic quantum well \cite{ref:Kim99}.  Pulsed excitation of semiconductor quantum dots (QD's) can also be used for
single-photon production \cite{ref:Michler00,ref:Santori01}. The energy of the photon emitted due to electron-hole recombination
in a dot depends on the total charge configuration of the dot \cite{ref:Landin98}. If we excite a QD with a laser pulse, then, the
electron-hole pairs that are created will each recombine to emit a photon with a unique wavelength.  A single emitted photon can
subsequently be isolated by spectral filtering \cite{ref:Gerard99}.

QD's offer several advantages as sources for single photons. They have high oscillator strengths and narrow spectral linewidths,
and do not suffer from photobleaching or shelving. The materials used to make QD's are compatible with mature semiconductor
technologies, allowing them to be further developed and integrated with other components.  A significant drawback, though, is that
very few of the photons emitted by a QD escape from the high-refractive-index semiconductor containing the dot into useful
directions. This can be remedied by placing the dot in a microscopic optical cavity, increasing the spontaneous emission rate by a
quantity known as the Purcell factor. The fraction $\beta$ of the emitted photons which are captured by the cavity mode then
depends on the enhanced emission rate $\gamma$ and the emission rate $\gamma_o$ in the absence of a cavity: $\beta = 1 - (\gamma_o
- \gamma_c)/\gamma$, where $\gamma_c / \gamma_o$ is the fraction of radiation that would be coupled into the cavity mode in the
limit of zero photon storage time.

In order to have an efficient source of single photons, it is necessary that a large fraction of the light escapes from the
confined cavity mode into a single travelling-wave mode.  This extraction efficiency can be determined by comparing the quality
factor $Q$ of the mode in a real cavity to the quality factor $Q_o$ for an ideal cavity without unwanted losses: $\eta_{\rm
extract} = Q/Q_{o}$. The mean photon number per pulse that can be observed will also depend on the total collection and detection
efficiency of the experimental apparatus. Since this is not intrinsic to the single-photon source, we will concentrate on the
external quantum efficiency $\eta$ of the device, independent of the measurement equipment.

Several semiconductor microcavities have been investigated for the enhancement of spontaneous emission from QD's, including
whispering-gallery modes in microdisks \cite{ref:Gayral99} and defect modes in two-dimensional photonic crystals
\cite{ref:Reese01,ref:Yoshie01}.  More practical for light extraction are microscopic posts etched out of distributed-Bragg
reflector (DBR) microcavities \cite{ref:Gerard98,ref:Solomon01}. Light escaping from the fundamental mode of a micropost
microcavity is well approximated by a Gaussian beam, and can thus be efficiently coupled into optical fibers, detectors, or other
downstream optical components.

We used molecular-beam epitaxy to grow planar DBR microcavities containing self-assembled InAs QD's.  The DBR mirrors consist of
alternating quarter-wavelength-thick layers of GaAs and AlAs, separated by a one-wavelength-thick spacer layer of GaAs.   The
reflectivity of the bottom DBR was deigned to be significantly higher than that of the top DBR, so that almost all of the light in
the cavity escapes upwards rather than downwards.  The QD's were grown at the center of the spacer layer. They are islands of InAs
formed by a strain-induced self-assembly process \cite{ref:Bimberg99}. We grew islands with a low areal density by using a high
substrate temperature and by stopping InAs deposition shortly after island formation.

\begin{figure}
\begin{center} \epsfig{file=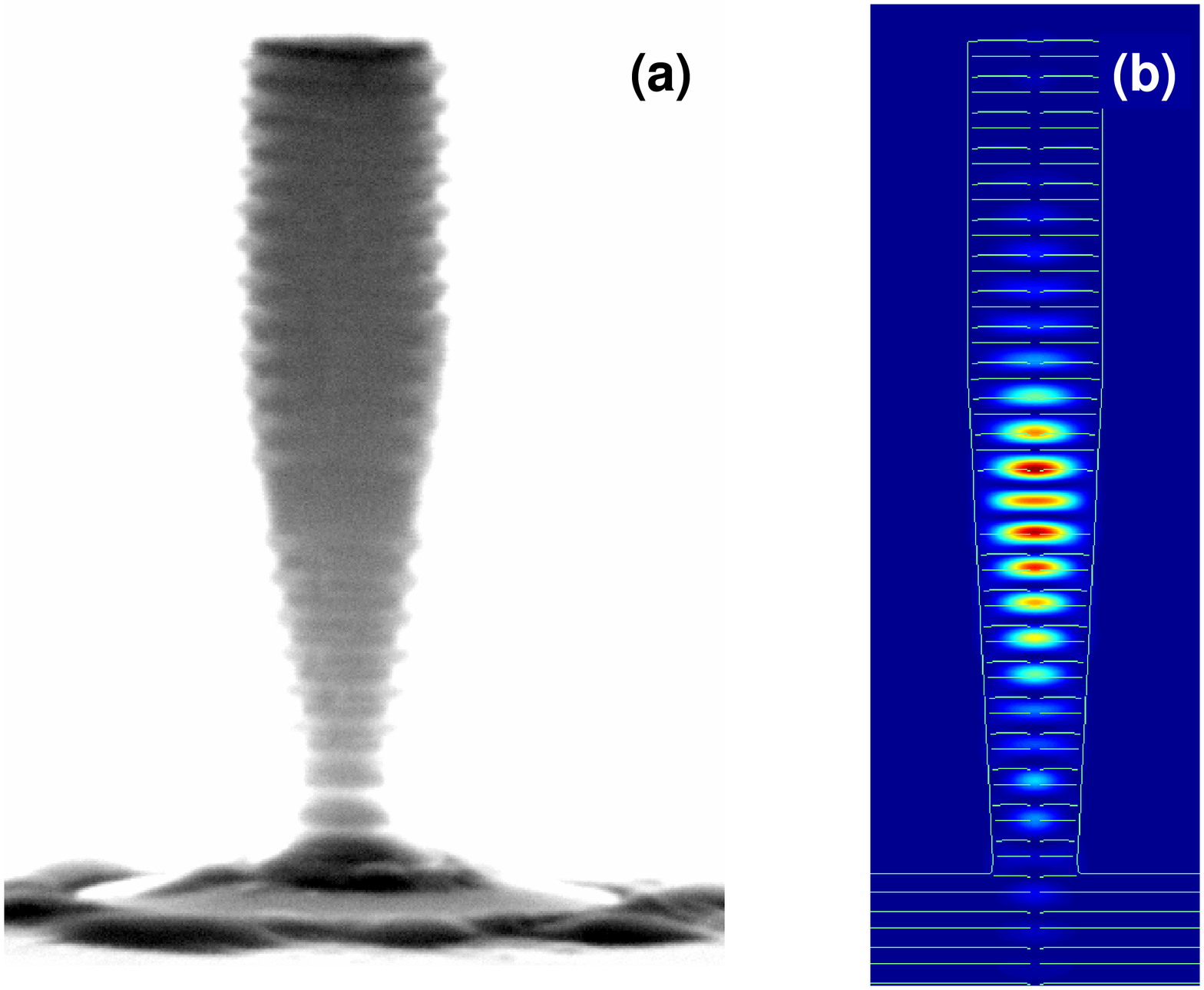,height=3in, angle=-90}
\end{center}
\caption{(a) Scanning-electron microscope (SEM) image of a micropost microcavity with a top diameter of 0.6 $\mu$m and a height of
4.2 $\mu$m. (b) Color-scale representation of the amplitude of the electric field for the fundamental mode of the micropost
microcavity, as calculated by the finite-difference time-domain method.  The profile of the modelled post matches the profile of
the real posts, as measured from SEM images.} \label{fig:new-post}
\end{figure}

Following the growth, we etched microposts out of the sample. A bilayer resist was exposed using an electron beam and was
subsequently used to lift off a thick nickel mask.  The sample was then etched using a low-pressure electron-cyclotron-resonance
plasma of chlorine and boron trichloride in a background of argon.  We divided the etch into three stages; in each subsequent
stage, we decreased the flow rate of chlorine and decreased the process pressure. The sample was cooled to an initial temperature
of about $3^o$C before the etch was started. Fig. \ref{fig:new-post}(a) shows a scanning-electron microscope image of a typical
etched micropost.  Light in the post is confined vertically by the DBR's and laterally by total internal reflection.

For optical measurements, pulsed laser light with a photon energy larger than the GaAs bandgap was directed towards the micropost.
The sample was held in a liquid-helium cryostat at a temperature of approximately 5 K, so that the created carriers were rapidly
trapped by the QD and quickly relaxed to the lowest-energy confined states. Optical emission was collected by a lens in front the
the cryostat and was filtered spectrally and spatially to eliminate scattered pump light. The emitted light could be sent to a
spectrometer (with a spectral resolution of 0.05 nm) or to a streak camera (with a temporal resolution of 25 ps), for measurement
of intensity as a function of time and of wavelength. Alternatively, it could be directed towards a Hanbury Brown and Twiss-type
(HBT) apparatus, which incorporated spectral filtering, in order to record a histogram of time intervals between photons. In the
limit of low total collection and detection efficiency, this histogram approximates the photon correlation function
$g^{(2)}(\tau)$. More detail on the experimental methods can be found in Ref. \cite{ref:Santori02c}.

We used selected a particular post, with a top diameter of 0.6 $\mu$m, which exhibited a single-QD photoluminescence line at a
wavelength of 855 nm, spectrally well removed from the wetting-layer emission.  A visibility of $33.1 \pm 1.8$ \% was measured in
a linear polarization basis for this emission line, while very low visibility was measured in a circular basis. We therefore
modelled light from the QD as consisting of a linearly polarized part together with an unpolarized part in determining the
fraction of QD emission lost at the polarizers in our HBT setup.

\begin{figure}
\begin{center} \epsfig{file=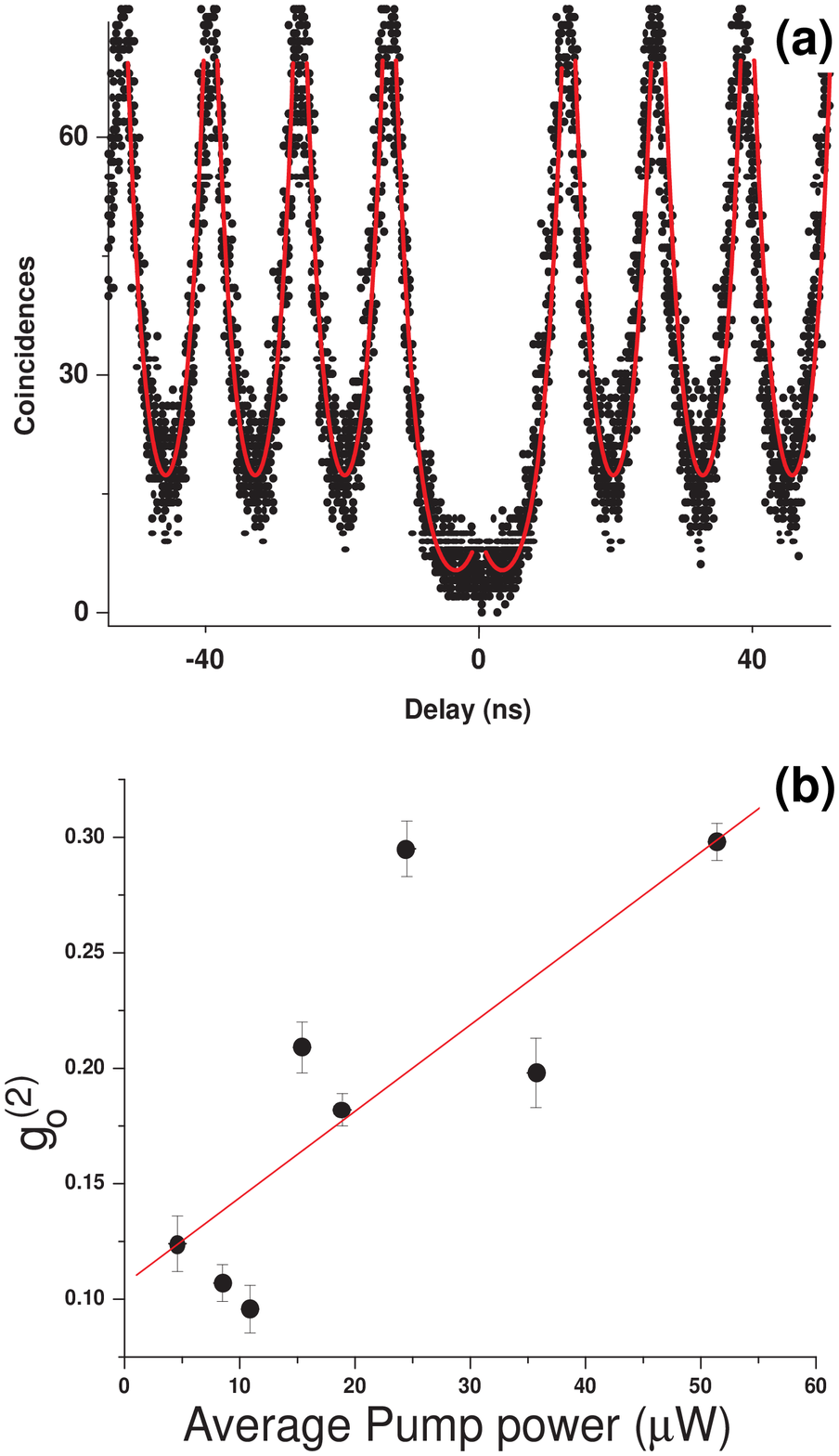,height=4in, angle=0}
\end{center}
\caption{(a) Measured autocorrelation for photons from a single quantum dot in a micropost microcavity, for an incident pump power
of 10.9 $\mu$W and an integration time of 300 s (points), and corresponding fit (line). Due to the relatively low emission rate
from the QD, adjacent peaks overlap. (b) Area of the central autocorrelation peak relative to the area of the side peaks as a
function of pump power (points).  The solid line is a guide to the eye.} \label{fig:g2-efficient}
\end{figure}

Single-photon generation was confirmed using the HBT apparatus. A measured histogram is shown in Fig. \ref{fig:g2-efficient}(a).
The central peak is nearly absent, reflecting strong suppression of the multi-photon probability. Each peak in the photon
correlation data can be described by a two-sided exponential, with a decay constant given by the spontaneous decay time and the
instrument response time. The recombination time was measured using the streak camera to be $4.4 \pm 1.2$ ns. (Details on lifetime
measurements are given in Ref. \cite{ref:Santori02}.) The time resolution of the HBT apparatus was calibrated by measuring
correlation functions for attenuated laser light scattered off the micropost. The width of the measured peaks was $473 \pm 29$ ps,
mostly limited by electronic jitter in our photon counters.

A fit using the measured time constants is also shown in Fig. \ref{fig:g2-efficient}(a);  the only two adjustable parameters are
the area of the central peak and the area of all the other peaks. The ratio of these areas, equal to $g_{o}^{(2)}$, is shown for
various pump powers in Fig. \ref{fig:g2-efficient}(b). The probability of multi-photon pulses increases with pump power,
suggesting that other states, apart from the desired QD emission, are contributing a background of unregulated photons.

The overall detection efficiency after the initial collection lens was determined by scattering attenuated laser light, tuned to
the QD emission wavelength, off the micropost of interest. The total photon count rate at the detectors was compared to the
optical power measured immediately after the collection lens using a calibrated power meter. Including light lost at the
polarizers, the detection efficiency was determined to be $3.02 \pm 0.16$ \%.

The fraction of light captured by the initial lens, on the other hand, was estimated to be 22\% using Gaussian-beam optics, with
the beam waist in the post being approximated by the fundamental mode in an infinite cylindrical waveguide. This estimate was
validated by calculations using the finite-difference time-domain (FDTD) method.  The field distribution in the fundamental mode
was calculated as described in Ref. \cite{ref:theory};  Fig. \ref{fig:new-post}(b) shows the result.  The far-field radiation
pattern was then estimated by Fourier-transforming the calculated near field \cite{ref:Vuckovic02}.  The result shows a nearly
Gaussian profile, with a divergence that agrees well with that given by Gaussian-beam optics.

In order to determine the device efficiency, the total photon count rate was normalized by the laser repetition rate and then
divided by the collection and detection efficiency, giving the mean photon number per pulse $\langle \hat{n} \rangle$.  We then
assumed that the light emitted from the source consists of a statistical mixture of perfectly regulated single photons, together
with a small background of photons with Poissonian statistics.  The coupling of this state into the travelling-wave mode leaving
the top of the micropost was modelled as an attenuation by a factor equal to the efficiency $\eta$, giving $\eta = \langle \hat{n}
\rangle (1-g^{(2)}_o)^{1/2}$.

\begin{figure}
\begin{center} \epsfig{file=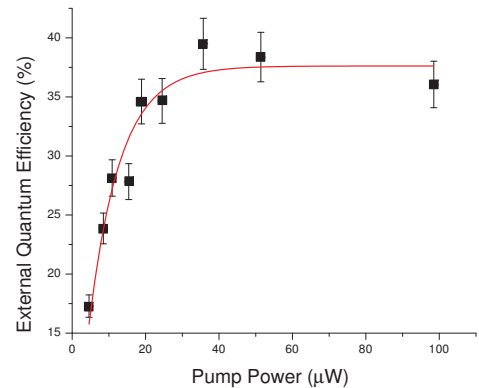,height=2in, angle=0}
\end{center}
\caption{External quantum efficiency of a single-photon source consisting of a single quantum dot in a micropost microcavity as a
function of pump power (points), together with a saturation fit (line).} \label{fig:efficiency}
\end{figure}

Fig. \ref{fig:efficiency} shows efficiencies determined in this way. The efficiency saturates at higher powers, when more than one
electron-hole pair is captured by the dot for each pump pulse. The solid line is a fit according to the saturation equation $\eta
= \eta_{\rm max} (1-e^{-P/P_{\rm sat}})$, where $P_{\rm sat}$ is the saturation pump power and the saturated efficiency $\eta_{\rm
max}$ is equal to $37.6 \pm 1.1$\%. This external quantum efficiency is approximately two orders of magnitude higher than that for
a QD in bulk GaAs.  We believe that this is also the highest efficiency yet reported for a single-photon source. We note that, in
determining this efficiency, we have not considered the polarization states of the emitted photons.  We also note that the
efficiency drops to approximately 8\% if we include losses at the initial collection lens; however, this could be improved simply
by using a lens with a larger numerical aperture.

The measured efficiency should be equal to the product of the coupling coefficient $\beta$ and the light extraction efficiency
$\eta_{\rm extract}$.  Under high pump power, the discrete QD transition is accompanied by a broadband background, filtered by the
cavity resonance; a Lorentzian fit to the filtered luminescence gives $Q = 628 \pm 69$. A similar measurement on an unetched
portion of the planar microcavity gives $Q_{o} = 1718 \pm 13$, resulting in $\eta_{\rm extract} = 36.6 \pm 4.0$\%.  The coupling
coefficient, on the other hand, was determined by measuring the recombination rates for other QD's on the same sample that are out
of resonance with the cavity mode.  The off-resonant lifetimes are expected to be nearly identical to the lifetimes of QD's in the
absence of a microcavity, due to the high density of leaky modes in our micropost microcavities \cite{ref:Bayer01}. Since the
unmodified lifetimes are too long to measure using our streak camera, we measured the autocorrelation of photons from these dots
using the HBT apparatus, and fitted the peak widths to obtain a recombination time of $25.4 \pm 1.4$ ns. This represents a Purcell
factor of $5.8 \pm 1.6$, corresponding to a coupling coefficient $\beta = 83 \pm 23$\%.  We note that the unmodified lifetimes are
unusually long for self-assembled InAs / GaAs QD's.  However, the demonstrated improvement of collection efficiency using the
Purcell effect is independent of the exact nature of the dots, and should apply equally to QD's with shorter lifetimes.

Combining the measured $\beta$ and $\eta_{\rm extract}$ results in an expected external quantum efficiency of $30 \pm 9$\%. This
agrees, within the error, with the efficiency that we determined at saturation.

Both $\beta$ and $\eta_{\rm extract}$ are limited by the quality factor of the microcavity mode. FDTD simulations predict $Q =
657$ for our micropost, in good agreement with the experimental value. Since the calculations do not include non-idealities such
as surface roughness, the difference between $Q$ for the micropost and $Q_o$ for the planar microcavity can be attributed to the
post geometry. Improving $Q$ requires more vertical post sidewalls, which should be achievable using different etching techniques,
such as chemically assisted ion-beam etching.  Optimization of the microcavity design can increase the quality factor yet further,
allowing for coupling efficiencies approaching 100\% \cite{ref:theory2}.

To summarize, we have demonstrated efficient generation of single photons using a single quantum dot in a micropost microcavity.
The emission rate from the dot was enhanced by a factor of 5.8, so that 83\% of the emitted light was coupled into a single cavity
mode.  The majority of this light escaped into a single-mode, Gaussian-like travelling wave, resulting in an external quantum
efficiency of approximately 38\%. This high efficiency is achieved at the same time that the probability of having more than one
photon in a given pulse is reduced by a factor of seven as compared to Poissonian light.  We note that single-photon generation
using a single QD in a micropost microcavity has recently been reported by other researchers \cite{ref:Moreau01b}, but no explicit
treatment of device efficiency has been provided.

An efficient source of single photons will be useful for quantum key distribution \cite{ref:Lutkenhaus00}.  Using existing
single-photon sources would result in a limited secure-key transmission rate over reasonable distances, due to the accumulated
effects of source inefficiency, channel loss, and compression during error correction and privacy amplification.  Our demonstrated
improvement in source efficiency would allow for transmission through approximately 20 dB of additional channel loss. Our
single-photon source may also eventually be useful for linear-optical quantum computation, providing the very high efficiencies
required, while emitting indistinguishable photons capable of exhibiting the necessary fourth-order interference
\cite{ref:Knill01}.

\begin{acknowledgments}
We would like to thank E. Waks and A. Scherer for helpful
discussions.  Financial assistance for M.P. and C.S. was provided
by Stanford University.  Financial assistance for C.S. was also
provided by the National Science Foundation. Financial assistance
for G.S.S. was provided by the Army Research Office.
\end{acknowledgments}

\bibliography{efficient}






\end{document}